\documentclass[aps,prl,twocolumn,superscriptaddress]{revtex4}
\usepackage{amsmath,amssymb}
\usepackage{graphicx}

\usepackage[usenames]{color}
\usepackage[bookmarks=true,colorlinks,linkcolor=red,urlcolor=blue,citecolor=blue]{hyperref}

\input epsf
\usepackage{epstopdf}

\definecolor{darkgreen}{rgb}{0,0.4,0}
\definecolor{darkred}{rgb}{0.4,0,0}
\definecolor{darkblue}{rgb}{0,0,0.4}

\def\be{\begin{equation}}
\def\ee{\end{equation}}
\newcommand{\bea}{\begin{eqnarray}}
\newcommand{\eea}{\end{eqnarray}}

\input epsf

\newlength{\extraspace}
\newlength{\extraspaces}
\def\bra#1{{\langle}#1|}
\def\ket#1{|#1\rangle}

\def\II{\relax{I\kern-.10em I}}

\def\IZ{\relax{\rm Z\kern-.34em Z}}
\def\IB{\relax{\rm I\kern-.18em B}}
\def\IC{{\relax\hbox{$\inbar\kern-.3em{\rm C}$}}}
\def\ID{\relax{\rm I\kern-.18em D}}
\def\IE{\relax{\rm I\kern-.18em E}}
\def\IF{\relax{\rm I\kern-.18em F}}
\def\IG{\relax\hbox{$\inbar\kern-.3em{\rm G}$}}
\def\IGa{\relax\hbox{${\rm I}\kern-.18em\Gamma$}}
\def\IH{\relax{\rm I\kern-.18em H}}
\def\II{\relax{\rm I\kern-.18em I}}
\def\IK{\relax{\rm I\kern-.18em K}}
\def\IP{\relax{\rm I\kern-.18em P}}

\def\inbar{\,\vrule height1.5ex width.4pt depth0pt}

\def\IR{\relax{\rm I\kern-.18em R}}

\def\lp10{\ell_p^{10}}
\def\lp11{\ell_p^{11}}
\def\R11{R_{11}}

\def\frac#1#2{{#1 \over #2}}

\newdimen\tableauside\tableauside=1.0ex
\newdimen\tableaurule\tableaurule=0.4pt
\newdimen\tableaustep
\def\phantomhrule#1{\hbox{\vbox to0pt{\hrule height\tableaurule width#1\vss}}}
\def\phantomvrule#1{\vbox{\hbox to0pt{\vrule width\tableaurule height#1\hss}}}
\def\sqr{\vbox{%
  \phantomhrule\tableaustep
  \hbox{\phantomvrule\tableaustep\kern\tableaustep\phantomvrule\tableaustep}%
  \hbox{\vbox{\phantomhrule\tableauside}\kern-\tableaurule}}}
\def\squares#1{\hbox{\count0=#1\noindent\loop\sqr
  \advance\count0 by-1 \ifnum\count0>0\repeat}}
\def\tableau#1{\vcenter{\offinterlineskip
  \tableaustep=\tableauside\advance\tableaustep by-\tableaurule
  \kern\normallineskip\hbox
    {\kern\normallineskip\vbox
      {\gettableau#1 0 }%
     \kern\normallineskip\kern\tableaurule}%
  \kern\normallineskip\kern\tableaurule}}
\def\gettableau#1 {\ifnum#1=0\let\next=\null\else
  \squares{#1}\let\next=\gettableau\fi\next}

 \def\eqnn#1{\xdef #1{(\secsym\the\meqno)}\writedef{#1\leftbracket#1}%
 \global\advance\meqno by1\wrlabeL#1}
 \def\eqna#1{\xdef #1##1{\hbox{$(\secsym\the\meqno##1)$}}
 \writedef{#1\numbersign1\leftbracket#1{\numbersign1}}%
 \global\advance\meqno by1\wrlabeL{#1$\{\}$}}
 \def\eqn#1#2{\xdef #1{(\secsym\the\meqno)}\writedef{#1\leftbracket#1}%
 \global\advance\meqno by1$$#2\eqno#1\eqlabeL#1$$}

\global\newcount\itemno \global\itemno=0

\def\itemaut#1{\global\advance\itemno by1\noindent\item{\the\itemno.}#1}

\def\({\left(}
\def\){\right)}

\usepackage{braket}
\usepackage[yyyymmdd,hhmmss]{datetime}
\newif{\ifeq}           
\eqtrue                 
\begin{document}

\title{Entanglement Constraints on States Locally Connected to the Greenberger-Horne-Zeilinger State}
\author{Grant W. Allen}
\affiliation{Department of Physics, University of California at San Diego, La Jolla, CA 92093-0354} 
\author{Orest Bucicovschi}
\affiliation{Department of Mathematics, University of California at San Diego, La Jolla, CA 92093-0112}
\author{David A. Meyer}
\affiliation{Department of Mathematics, University of California at San Diego, La Jolla, CA 92093-0112}

\begin{abstract}
The multi-qubit GHZ state possesses tangles with elegant transformation properties under stochastic local operations and classical communication. Since almost all pure 3-qubit states are connected to the GHZ state via SLOCC, we derive a necessary and sufficient achievability inequality on arbitrary 3-qubit tangles, which is a strictly stronger constraint than both the monogamy inequality and the marginal eigenvalue inequality. We then show that entanglement shared with any single party in the $n$-qubit GHZ SLOCC equivalence class is precisely accounted for by the sum of its $k$-tangles, recently coined the strong monogamy equality, acknowledging competing but agreeing definitions of the $k$-tangle on this class, one of which is then computable for arbitrary mixed states. Strong monogamy is known to not hold arbitrarily, and so we introduce a unifying outlook on entanglement constraints in light of basic real algebraic geometry.
\end{abstract}
\maketitle
%


The physical presence of entanglement permits access to and manipulation of various quantum mechanical information without the physical presence of each subsystem \cite{w13}. This remarkable privilege appears to come at a price however --- as entanglement is found to be partner preferential, the postmodern polyamorist subculture is marginalized \cite{CKW00, OV06, SAPb12, OCF14, KGS, OF07}, that is to say, the sharing of entanglement is limited by physical law. Therefore, a common milestone in quantum technology is sufficient control of substrates for the generation of highly multi-partite highly entangled states, for example, the Greenberger-Horne-Zeilinger state, $\ket{\text{GHZ}} = \ket{0}^{\otimes n}+\ket{1}^{\otimes n}$. This state alone has already garnered significant attention, providing stronger tests against local realism \cite{GHZ89}, improving precision atomic clocks \cite{KKBJSYL} and interferometry \cite{BIWH96} with possible implications for gravitational wave detection \cite{Metal17}, thereby further motivating recipes for GHZ generation \cite{FFPP11, CZDP17, GCD09} with record setting confirmations in the number of qubits on varying hardware \cite{M11, Leib05, Gao10, Yao12}.

Once such a highly entangled state is obtained, the next cheapest layer of information processing consists of applying (stochastic) local operations and classical communications (SLOCC). Based on the presumed relative abundance of SLOCC implementation, the local operations serve another purpose --- for the construction of a value to assign to states which lie outside SLOCC's own ability to generate, \textit{i.e.}, non-local states \cite{G17}. The value, as often the case, will simply be referred to as an amount of entanglement, ignoring potentially more fine-grained notions of the non-local resource. 

Consider that if two states can be probabilistically transformed back \textit{and} forth with SLOCC, the states should contain, in some sense, the same non-local resources, so that entanglement can be thought of as \textit{invertible}-SLOCC invariant.  In a SLOCC transformation, state norms only correspond to the success probability and are otherwise irrelevant, so it is convenient to identify invertible SLOCC operations with the classic unit determinant group $SL(2,\mathbb{C})^{\otimes n}$. It just so happens that any homogeneous $SL$ invariant is guaranteed to be non-increasing on average under complete combinations of local operations, and thus meets the standard prerequisite of entanglement monotonicity \cite{V00, VDD03, EBOS12}. As we will see, it is convenient and aesthetically pleasing to choose the generators of the algebra of $SL$ invariants to be the relevant entanglement measures. The algebra of invariants for $n$ copies of $SL(2,\mathbb{C})$ is partially known \cite{GW13} and can be generated by the determinant for $n=2$ \cite{P11}, and the hyperdeterminant for $n=3$ \cite{LTT07}, which are often normalized to give the pure 2-tangle,  $\tau_{A|B}(\psi_{i,j}) = 2 |\text{det}(\psi_{i,j}) |$ and the pure 3-tangle, $\tau_{A|B|C}(\psi_{i,j,k}) = 2 \sqrt{| \text{hdet}(\psi_{i,j,k} )| }$ where $\psi$ is the pure state tensor coefficients and the extra root on the 3-tangle ensures for both tangles identical transformation properties under local $GL$ operations \cite{VDD01, T13}. Both tangles are extended to mixed states linearly on the particular pure state decomposition which provides the minimal average tangle, known as the convex roof \cite{W98, O16}.

The celebrated results  of D\"ur \textit{et al.}~\cite{DVC00} show that a generic 3-qubit pure state can be transformed to the GHZ state with SLOCC, meaning that a random sample from Hilbert space will almost always produce a state connected to GHZ. By calculating tangles of the state $M_1\otimes M_2\otimes M_3 \ket{\text{GHZ}}$ with $M_i$ any $2\times2$ complex matrix, which reaches almost all states, we derive a necessary and sufficient inequality to describe all possible \textit{values} of tangles in three qubits, and we find this inequality to be strictly stronger than both monogamy \cite{CKW00} and marginal eigenvalue \cite{HSS03} inequalities. The crux of our results relies on the fact that the SLOCC invariants are \textit{not} invariant when SLOCC acts externally to the relevant parties, in which case, we find the corresponding transformation rules simplify elegantly once restricted to the GHZ class; this Letter is about exploring the implications. 

For more qubits, the GHZ class loses its generality, but we show that it retains a remarkable monogamy property as follows. Recall for arbitrary 3-qubit states, the total entanglement with any party $A$, $\tau_A = 2\sqrt{\text{det}\rho_A}$, also known as the 1-tangle, decomposes exactly in terms of tangles with other parties $B$ and $C$ \cite{CKW00},
\begin{equation}\label{3exact}
\tau_A^2 = \tau_{A|B}^2+\tau_{A|C}^2+\tau_{A|B|C}^2.
\end{equation}
Recently, variations of the above have been conjectured to hold for all $k$-tangles in $n$-qubits, known as the strong monogamy relation \cite{RLMA} --- the most natural generalization for $n$ pure qubits is, 
\begin{equation}\label{mainresult}
\tau_{A}^2  \overset{?}{=} \sum_{\mathcal{I}_A} \tau_{\mathcal{I}_A}^2,
\end{equation}
where $\mathcal{I}_A$ is any subset of the $n$ parties that includes the party $A$ with $|\mathcal{I}_A|\geq2$. While strong monogamy \textit{as such}, appears to be a stretch for arbitrary states \cite{ROA16}, we find that it holds exactly on the GHZ SLOCC class. Thus we have excavated the very origin of the first law, Eq.~\ref{3exact}, that effectively seeded the entire field of distributed entanglement theory. And finally, in salvation to the failure of Eq.~\ref{mainresult} on all states, we describe a universal framework of entanglement constraints, in turn unearthing the origin of their existence as well.

Starting with three qubits, we calculate all of the tangles of the state, $M_1\otimes M_2\otimes M_3 \ket{\text{GHZ}}$. While it is a straightforward calculation, we lay some guideposts in the supplementary materials for any uncommon maneuvers \cite{ABMsup}. We find that the complex parametrization of the $M_i$s are hugely redundant and the tangle expressions only depend explicitly on four real parameters, $(r, \phi_1, \phi_2, \phi_3)$,
\begin{equation}
\begin{alignedat}{2}
\label{xyzt}
\tau_{B|C} &= \lambda c_1 s_2 s_3, \text{\space \space \space \space \space} \tau_{A|C} && =  \lambda s_1 c_2 s_3, \\
\tau_{A|B} &=  \lambda s_1 s_2 c_3, \text{\space \space }  \tau_{A|B|C} &&=  \lambda s_1 s_2 s_3,
\end{alignedat}
\end{equation}
where $c_i, s_i$ = (cos)sin$(\phi_i)$, $\phi_i\in[0,\pi/2]$, $\lambda=1/(r- c_1 c_2 c_3)$, and $r\geq1$, noting the pleasing and perhaps surprising symmetry sparkling amongst the collection. This is a remarkable simplification because we have 4 different tangles and only 4 parameters. We can find constraints on the tangles by inverting the expressions, Eq.~\ref{xyzt}, using standard algebraic tools, leading to the following theorem. Note that the full mathematical interruption is given in the supplementary materials \cite{ABMsup}. 

{\bf Achievability Theorem:} \textit{Given an arbitrary 3-qubit pure state $\ket{ \psi_{ABC}}$, the corresponding tangles, $(\tau_{B|C},\tau_{A|C},\tau_{A|B},\tau_{A|B|C})\equiv (x,y,z,t) $, satisfy the following inequality, }
\begin{equation}\label{xyztineq}
t^2 ( 1-x^2-y^2-z^2-t^2)-( x^2 y^2 + x^2 z^2 +y^2 z^2 - 2 x y z) \geq 0. 
\end{equation}
\textit{Conversely, for any non-negative 4-tuple $(x,y,z,t)$ satisfying the inequality, there exists a pure 3-qubit state with corresponding tangles. }

\begin{figure}[h]
	\centering
		\includegraphics[width=1in]{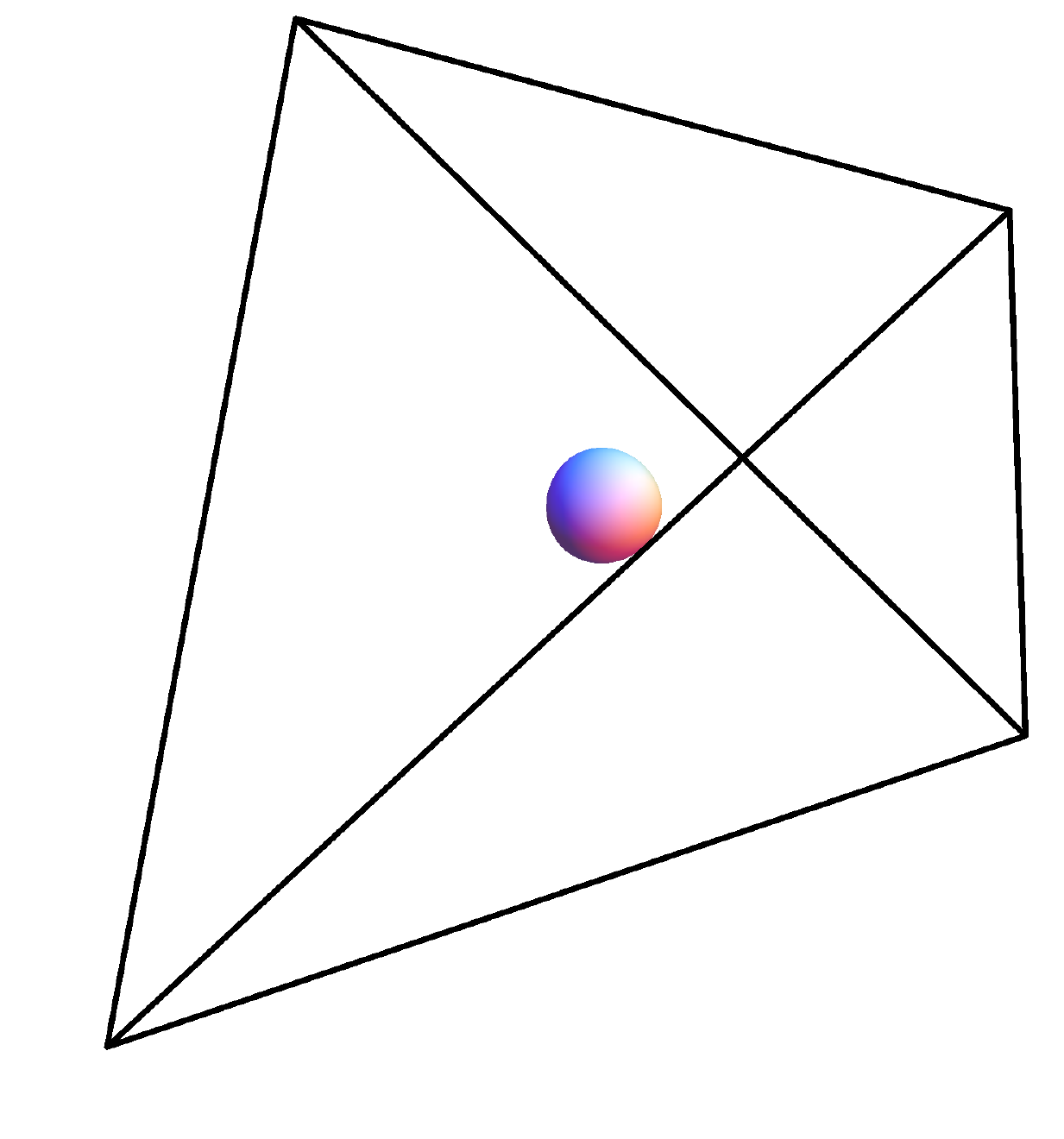}
		\includegraphics[width=1in]{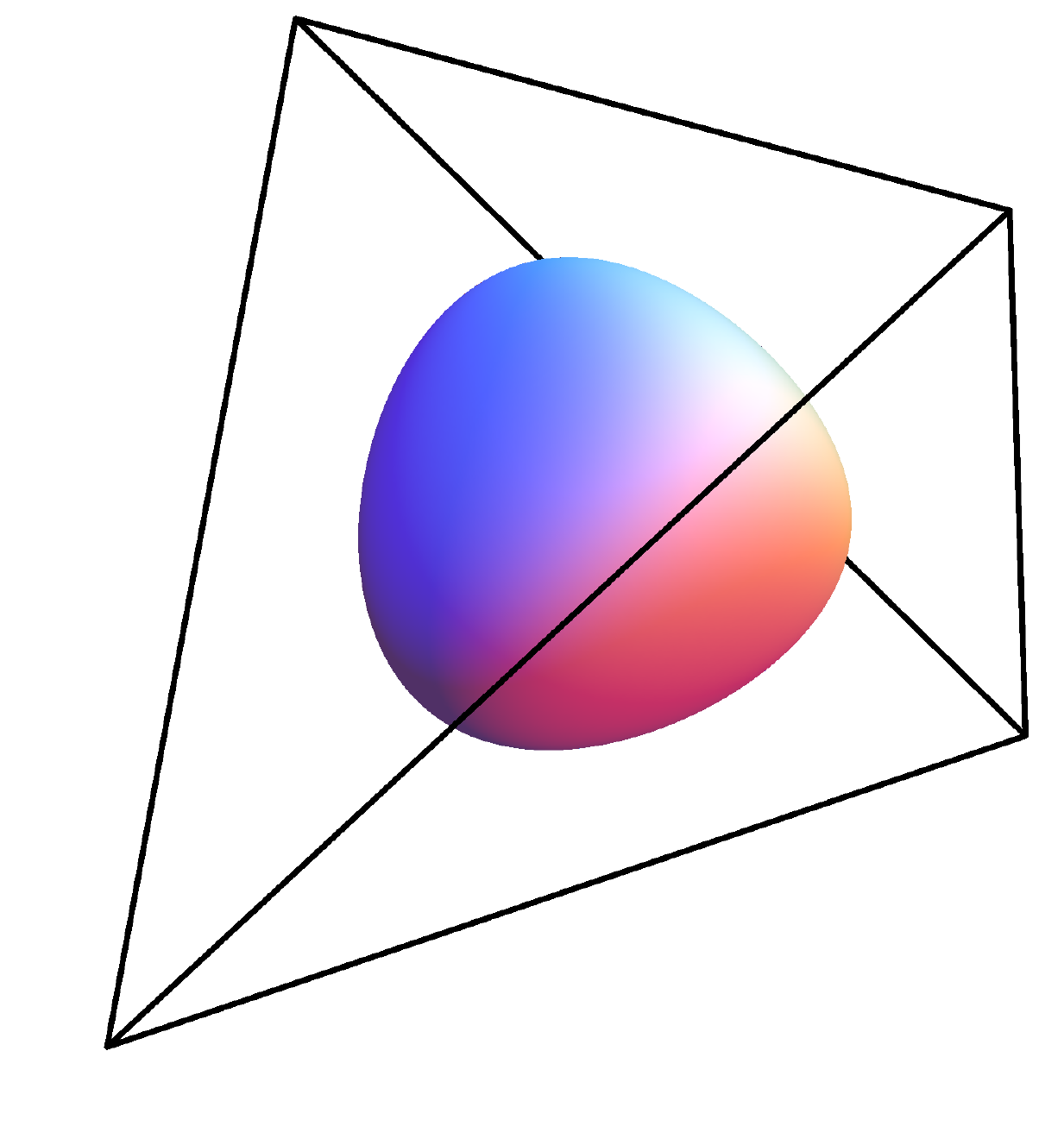}
		\includegraphics[width=1in]{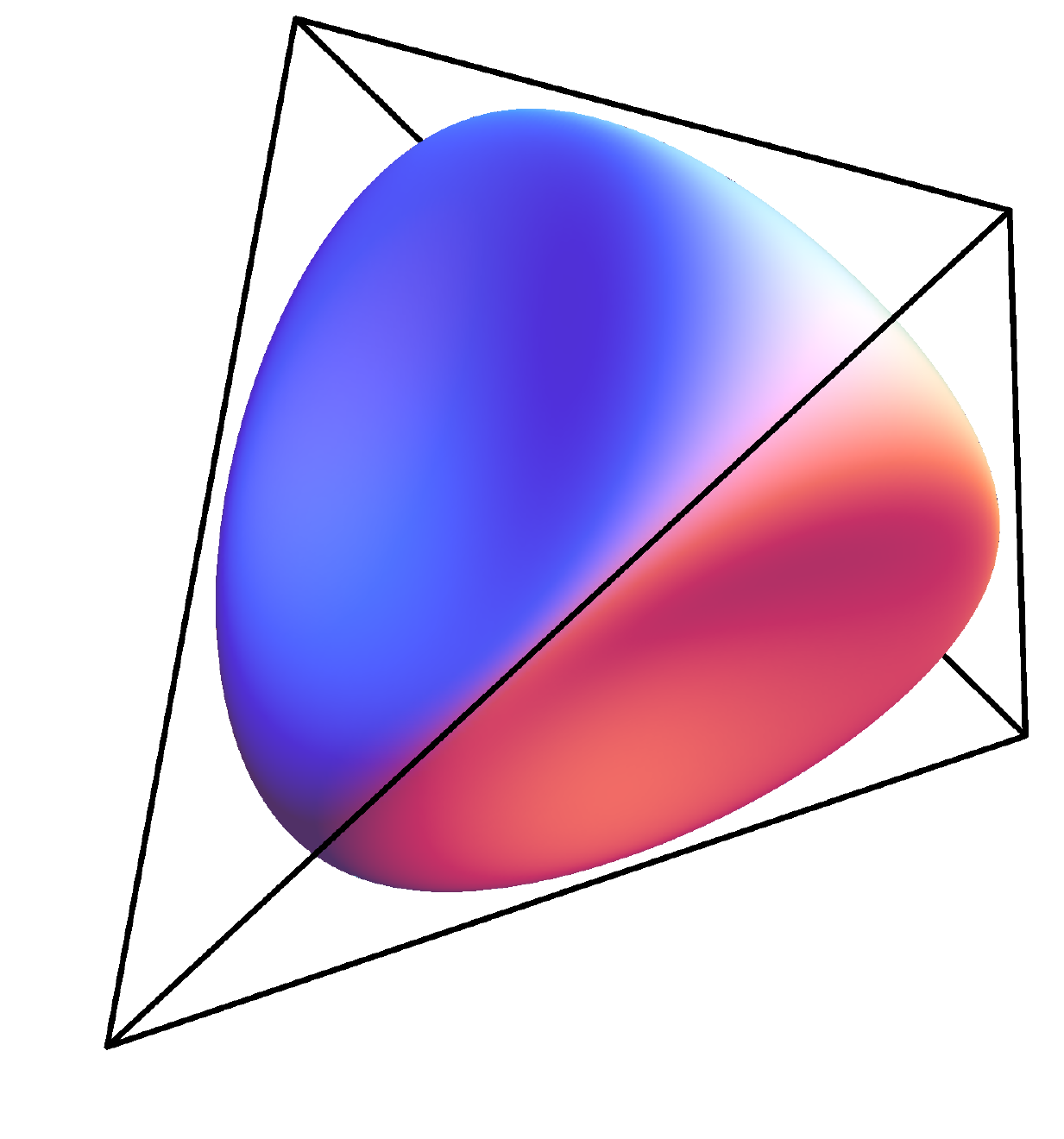}
		\includegraphics[width=1in]{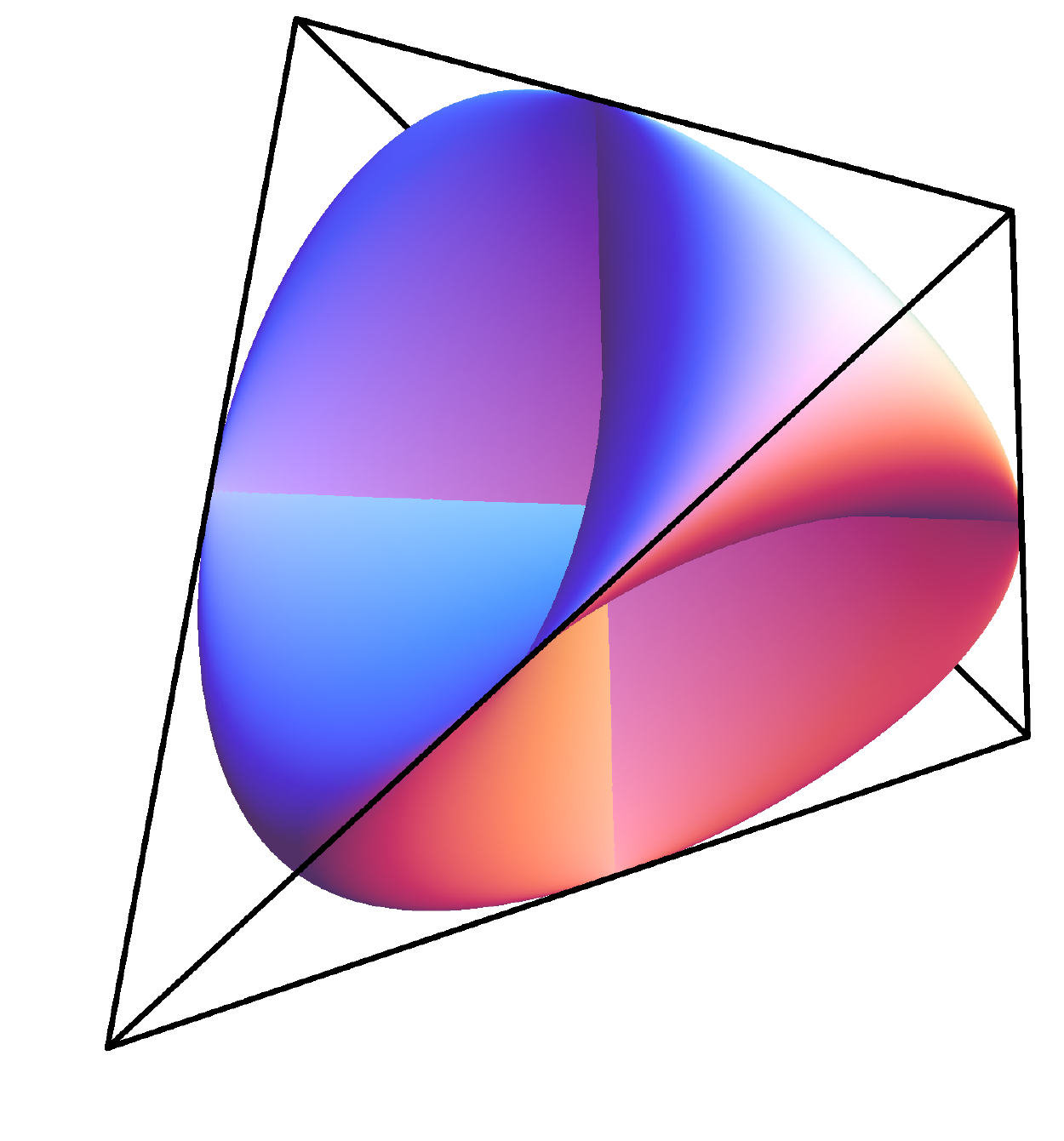}
		\includegraphics[width=1in]{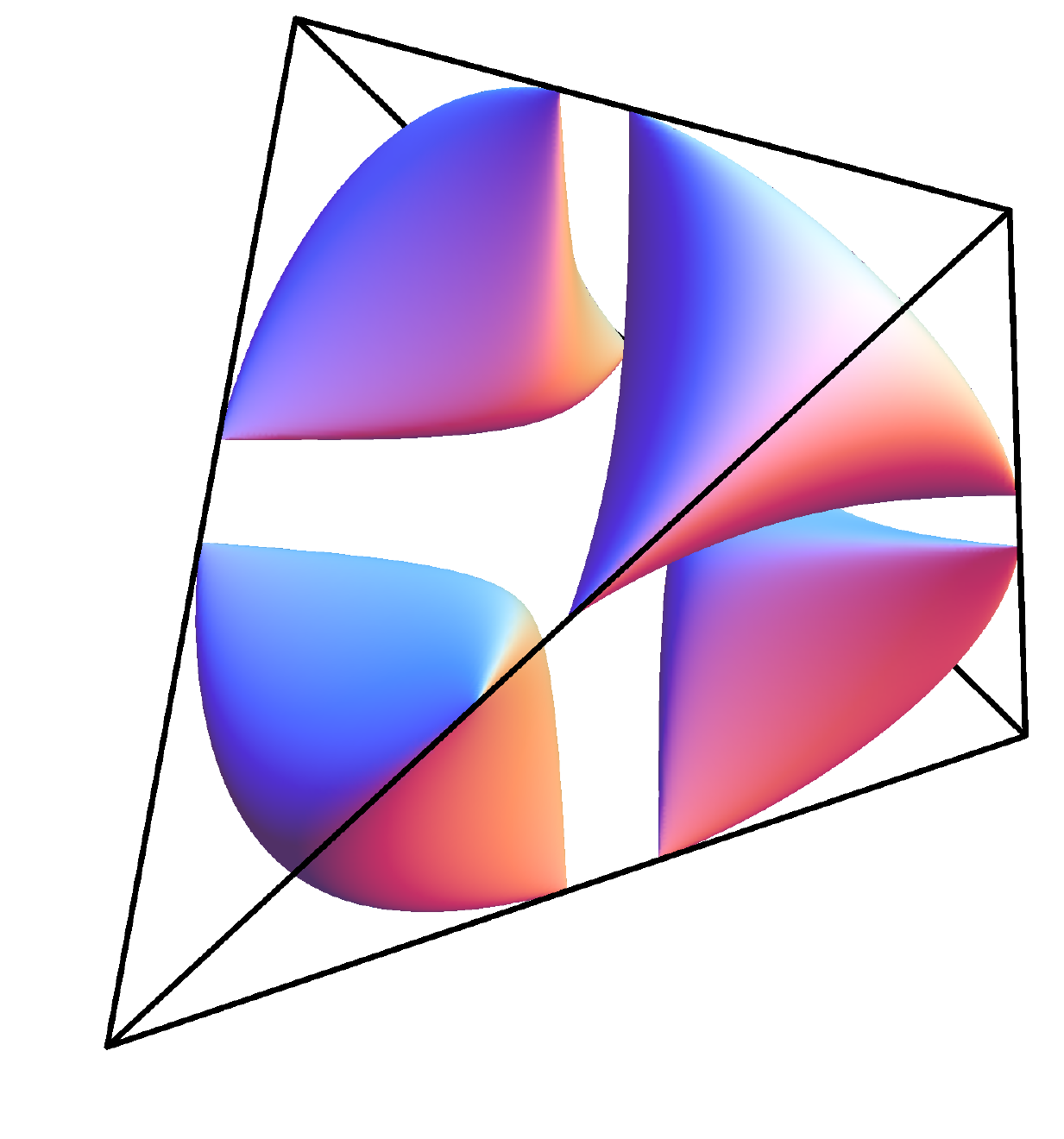}
		\includegraphics[width=1in]{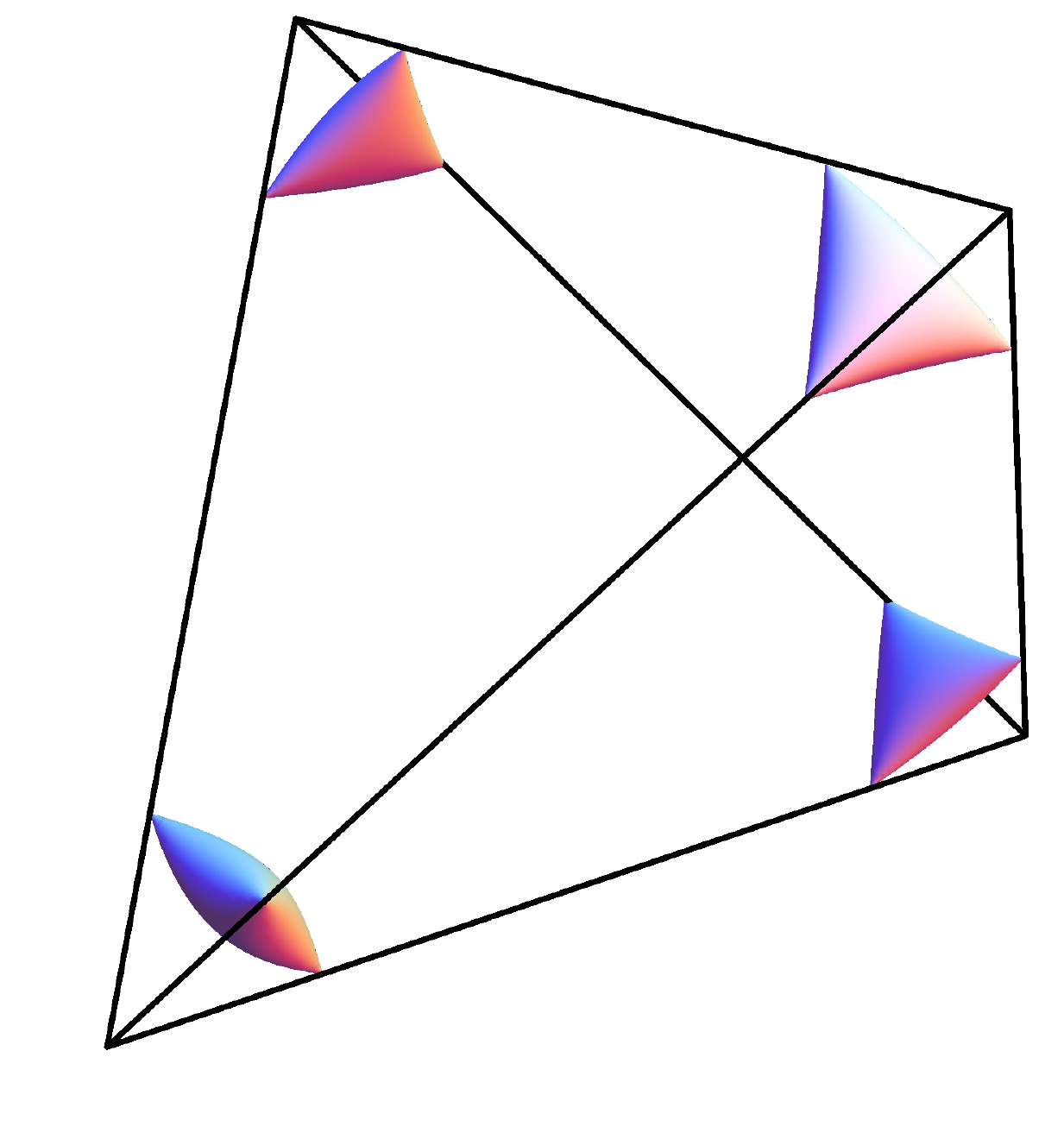}
	\caption{Envelope of slices to solution set of Eq. \ref{xyztineq} for $t^2= (.98,.64,.09,0,-.01, -.25)$}
	\label{cct}
\end{figure}

See Fig.~\ref{cct} for the solution set of Eq. \ref{xyztineq}, over the extended region, $ [ -1,1 ]^3$ in the $(x,y,z)$-subspace, for various values of $t^2 \in [-1,1]$; the black straight lines form the wire frame of the tetrahedral envelope. Note that an imaginary value for the 3-tangle is clearly non-sensical, however we soon find that corresponding surfaces are physically relevant. 

With the above achievable set of tangles being described by a single inequality, it is straightforward to project out the 3-tangle, \textit{i.e.,} take the union of $0\leq t \leq 1$ slices, and get a constraint on 2-tangles alone. It turns out this set can also be described by a single inequality. By noticing that Eq. \ref{xyztineq} is quadratic in $t^2$, we complete the square,
\begin{align}
\begin{split}\label{discq}
(1 -& 2 t^2 - x^2 - y^2 - z^2)^2 \\
\leq& \, (1 + x + y + z) (1 + x - y - z) \\
&\times(1 - x + y - z) (1 - x - y + z) ,
\end{split}
\end{align}
and we take the root, setting $t=0$ to unconstrain $(x,y,z)$ as much as possible, which gives the necessary and sufficient 2-tangle achievability inequality, reproducing the results of \cite{BGM}, and summarized further below.

It is often glossed over that the 2-tangles are defined as the minimum over all possible averages in a mixed state (the convex roof construction). The maximum average can also reveal useful information. Known through the \textit{concave} roof construction \cite{U}, the maximum average tangle in mixed states is commonly referred to as the tangle of assistance. The 2-tangle of assistance is not an entanglement monotone on 2-qubit states, however, it is an entanglement monotone on 3-qubits when evaluated among any 2-qubit pair \cite{GMS05} and may be related in a simple way to violations of Mermin inequalities \cite{CJKLL}. Since the tangle of assistance gives information about 3-qubit entanglement, one expects it to be related to the 3-tangle. Indeed, the following relation holds for any 3-qubit pure state, $\ket{\psi_{ABC}}$:
\begin{equation}\label{tauassistance}
\tau^2_{A|B|C} = \hat{\tau}^2_{A|B}-\check{\tau}^2_{A|B}= \hat{\tau}^2_{A|C}-\check{\tau}^2_{A|C}= \hat{\tau}^2_{B|C}-\check{\tau}^2_{B|C},
\end{equation}
where $\hat{\tau}$ denotes the concave roof tangle, and $\check{\tau}$ denotes the convex roof tangle (being the same tangle from the achievability theorem), so that the 3-tangle is the difference between the maximal and minimal 2-tangle among any pair. When the above is rearranged and substituted into Eq.~\ref{xyztineq}, one gets a necessary and sufficient 2-tangle of assistance inequality, however, peculiarly, if we examine the boundary, by turning the inequality into an equality, and squaring away the square root that appears, then factoring the result, we get the following polynomial as a factor,
\begin{equation}
-t^2(1-x^2 - y^2 - z^2 + t^2) - (x^2 y^2 + x^2 y^2 + y^2 z^2 -  2 x y z)=0,
\end{equation}
where now $(\hat{\tau}_{A|B},\hat{\tau}_{A|C},\hat{\tau}_{B|C}, \tau_{ABC}) \equiv (x,y,z, t)$. Note the curious relation to Eq.~\ref{xyztineq} by a Wick-like rotation, $t \rightarrow i t$. Revisit Fig. \ref{cct} for the solution set of Eq. \ref{xyztineq} for various values of $t^2 \in [-1,1]$. One can take the union of imaginary $t$ slices by a similar method as the real $t$ slices, which we summarize in the following theorem which includes both convex and concave roof cases distinguished with parenthesis and the symbol $\pm$ respectively. 

{\bf Corollary:} \textit{Given an arbitrary 3-qubit pure state $\ket{ \psi_{ABC}}$, its pairwise convex (concave) roof 2-tangles, $(\tau_{A|B},\tau_{A|C},\tau_{B|C}) \equiv (x,y,z)$, satisfy the following inequality,}
\begin{align}
\begin{split}\label{ccoaineq}
&\sqrt{(1-x-y+z)(1-x+y-z)} \\
& \,\,\,\,\,\,\overline{\times(1+x-y-z)(1+x+y+z)} \\
&\pm (1-x^2-y^2 - z^2 )\geq 0.
\end{split}
\end{align}
\textit{Conversely, for any non-negative triple $(x,y,z)$ satisfying the inequality, there exists a pure 3-qubit state with corresponding convex (concave) roof 2-tangles. }\

\begin{figure}[h]
	\centering
		\includegraphics[width=1in]{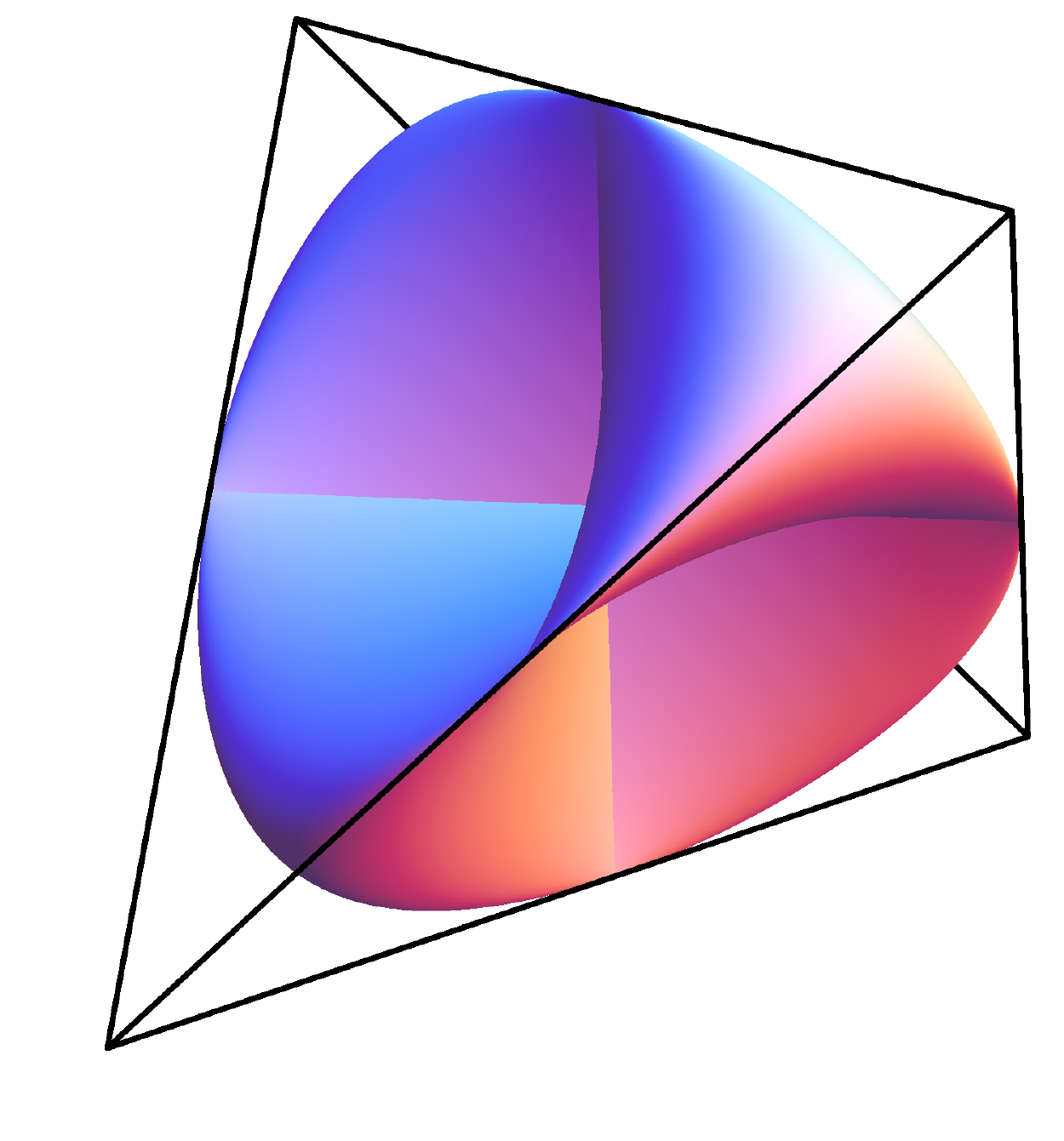}
		\includegraphics[width=1in]{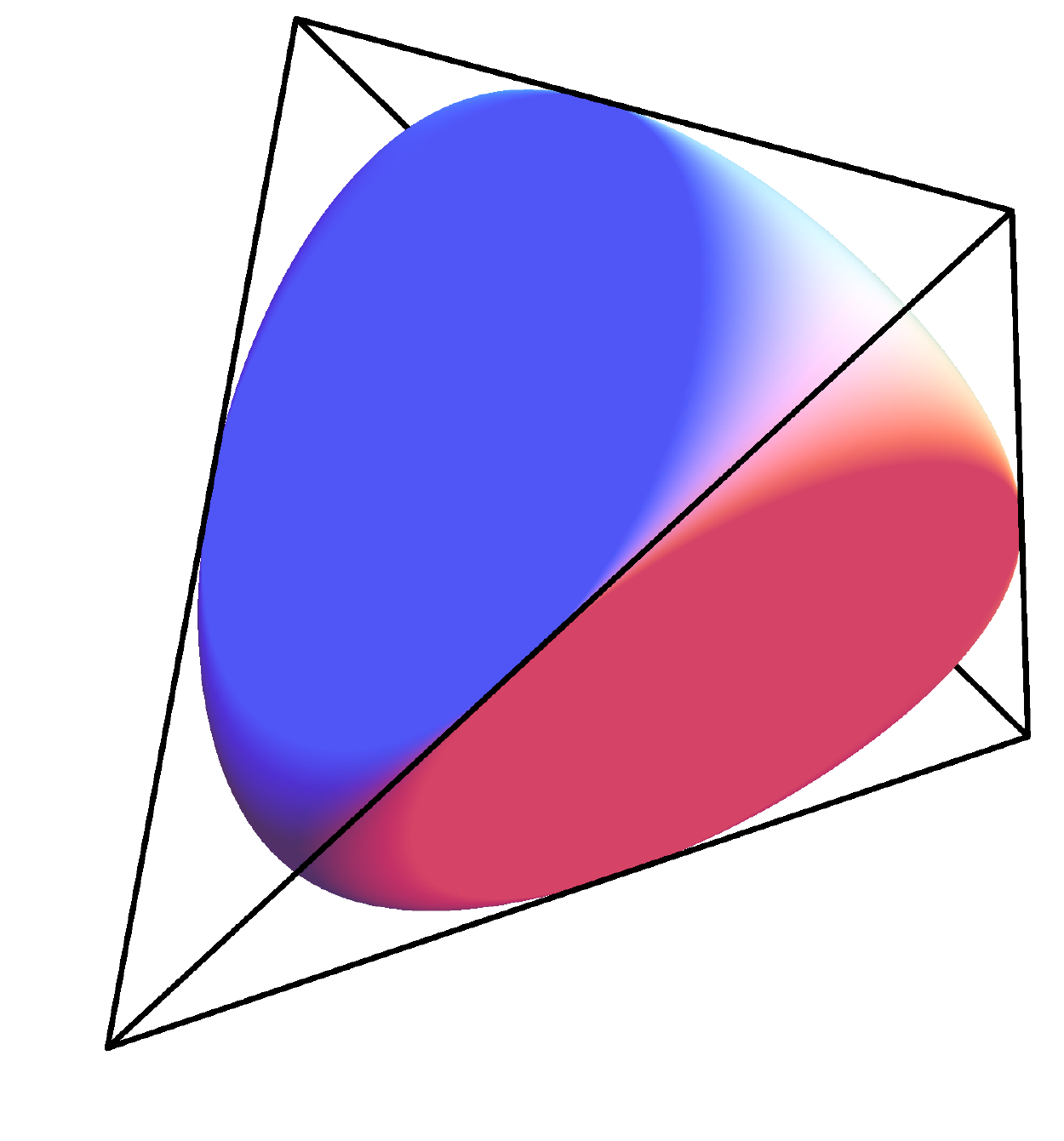}
		\includegraphics[width=1in]{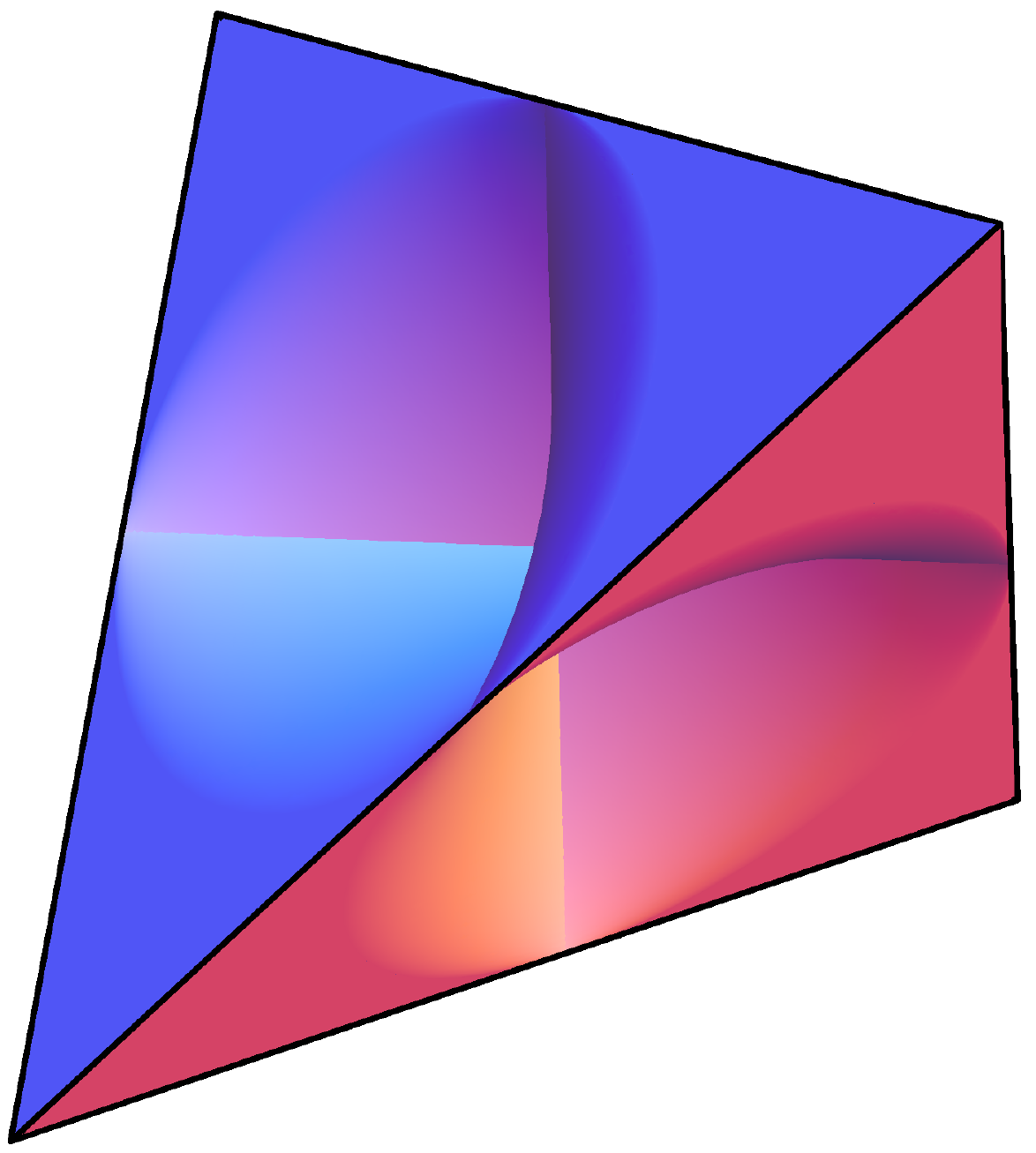}
	\caption{The Steiner Suite. Achievable convex (concave) roof 2-tangles form the non-negative component of the Steiner surface's (inverted) convex hull. The Steiner volume on the left can be thought of as coming from states in the null cone $\tau_{A|B|C}=0$ whose mixed 2-tangles are independent of the decomposition $\hat{\tau}_{i|j}=\check{\tau}_{i|j}$}
	\label{CCfig}
\end{figure}

The solution set of the inequalities are variations on the famous Roman Steiner surface, see Fig. \ref{CCfig} where the parenthesis again indicates separate cases. 

The convex hull of the surface has also shown up in a number of places, \textit{e.g.,} the parametrization of tripartite Werner states where 2-party marginals are constrained to be local-positive \cite{JV15}, in the image set of a triple of hermitian matrices \cite{SWZ16}, and finally in a quite peculiar classical/quantum duality \cite{BZ13, BGM, BDM}.

The story so far is similar in spirit to the marginal problem - given subsystem information, what are the consistency conditions for a joint state? Recall Eq.~\ref{3exact}, and that the 1-tangle can be written in terms of the minimal single party eigenvalue, $\lambda_A$, $\tau_A = 2\sqrt{\lambda_A (1-\lambda_A)}$, so we can relate the eigenvalues and invariants,
\begin{align}
\begin{split}
\lambda_A &= \frac{1}{2}\left( 1-\sqrt{1-\tau_{A|B}^2 - \tau_{A|C}^2-\tau_{A|B|C}^2} \right).
\end{split}
\end{align}

\begin{figure}[h]
	\centering
		\includegraphics[width=1.3in]{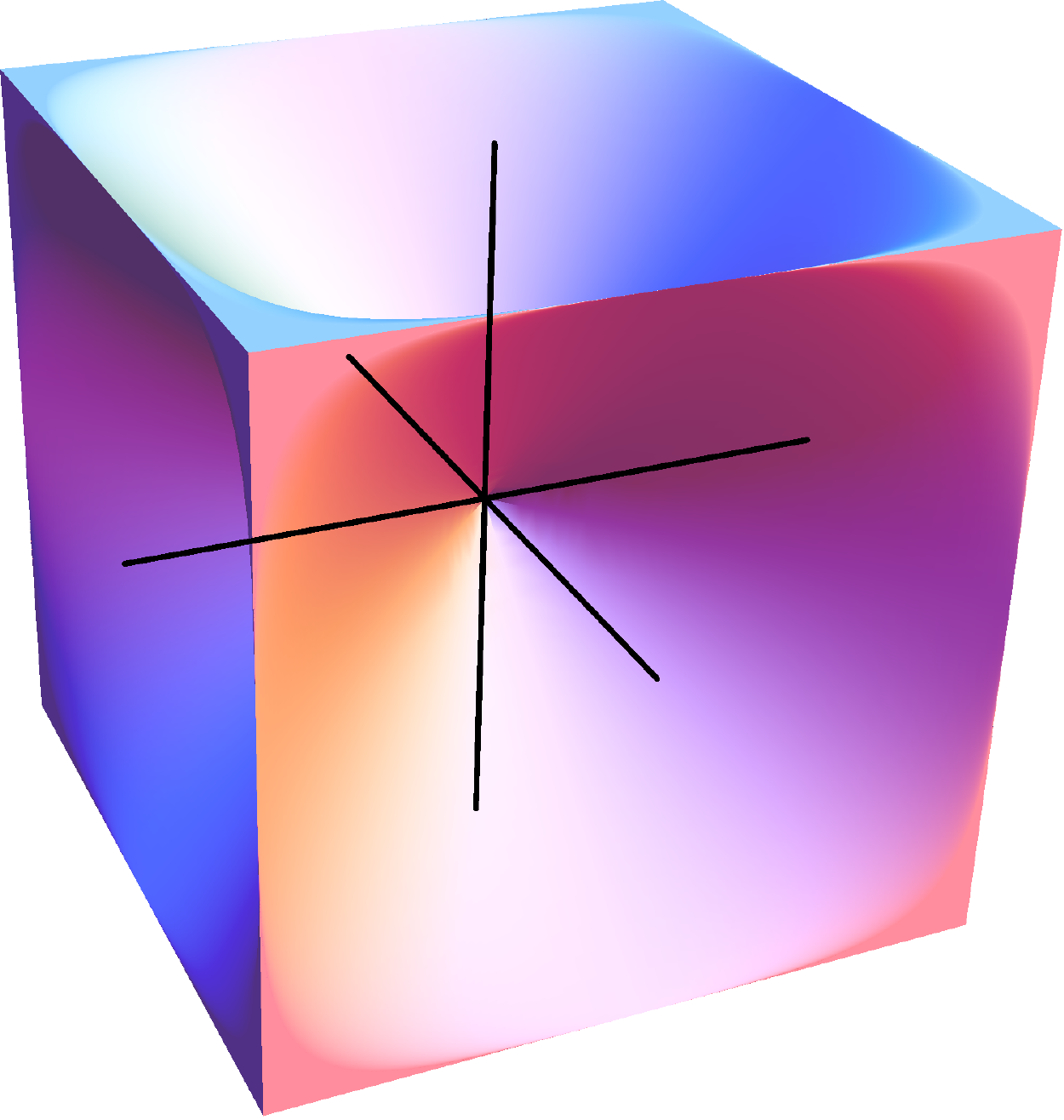}
	\caption{Achievable 1-tangles in 3-qubits form the non-negative component of the pictured volume. Note, the faces of the cube have been bored to the origin.}
	\label{1Tfig}
\end{figure}

By inverting the above with its analagous expressions for $\lambda_B$ and $\lambda_C$, one can rewrite Eq.~\ref{xyztineq} in terms of the eigenvalues to get a necessary and sufficient inequality on $(\lambda_A,\lambda_B,\lambda_C,\tau_{A|B|C})$. The 3-tangle can be projected out to recover the marginal eigenvalue inequality,
\begin{equation}
(\lambda_A-\lambda_B-\lambda_C)(-\lambda_A+\lambda_B-\lambda_C)(-\lambda_A-\lambda_B+\lambda_C)\geq 0,
\end{equation}
showing that Eq.\ref{xyztineq} is a strictly stronger constraint, where more detail is in the supplementary materials \cite{ABMsup}.
It is worth pointing out that one can rewrite this expression back in terms of the 1-tangles with the substitution, $\lambda_i = \frac{1}{2}(1-\sqrt{1-\tau_i^2})$ and further, the above marginal inequality has been extended to arbitrary numbers of pure qubits \cite{HSS03}, so the distribution of 1-tangles is hence likewise a fully solved problem. See Fig.~\ref{1Tfig} to see what the set of triples $(\tau_A, \tau_B, \tau_C)$ looks like in 3-qubits.

So far everything discussed follows readily from the tangle expressions of Eq.~\ref{xyzt}, yet the obvious symmetry still seems to be begging to be let further out of the box --- let us preemptively generalize to arbitrary numbers of qubits for our main result, writing down the expressions for the tangles, and afterwards discuss how we can derive the expressions in two independent ways.

{\bf Proposition:} \textit{Let $\mathcal{P}$ be a set of $n$ parties. The $k$-tangle between a subset of parties $\mathcal{I} \subseteq \mathcal{P}$ with $|\mathcal{I}|=k\geq 2$, of the $n$-qubit state, $\ket{\psi} = \otimes_{p\in \mathcal{P}} M_p \ket{\text{\normalfont{GHZ}}}$, is given by,}
\begin{equation}
\tau_{\mathcal{I}} = \frac{1}{r- \prod_{p \in \mathcal{P} } c_p }  \prod_{i \in \mathcal{I}} s_i \prod_{\bar{\imath} \in \bar{\mathcal{I}} } c_{\bar{\imath}},
\end{equation}
\textit{and the 1-tangle, for any single party $A$, is given by,}
\begin{equation}
\tau_{A} =  \frac{1}{r- \prod_{p\in \mathcal{P}} c_p } s_A\sqrt{1-\prod_{i \in \mathcal{P}\setminus \{A\}} c_i^2},
\end{equation}
\textit{where each $M_p\in \mathbb{C}^{2 \times 2}$, and without loss of generality, we choose a non-redundant real parametrization $M_p = \begin{pmatrix} u_p & v_p c_p \\ 0&  v_p s_p \end{pmatrix}$ with $u_p, v_p \geq 0$ and $c_p, s_p = ({\text{\normalfont cos}})\sin(\phi_p)$ and $2r = \prod_{p\in\mathcal{P}}\frac{u_p}{v_p}+\prod_{p\in\mathcal{P}}\frac{v_p}{u_p}$.}

Note that Eq.~\ref{mainresult} is satisfied by these tangles, which follows from a trivial trigonometric identitiy, $\prod_k (s_k^2 + c_k^2) = 1$. Therefore, this is the first family of states that satisfy the strong monogamy equality with all $k$-tangles being non-zero \cite{K14}. 

We now justify our formula for the $k$-tangle in two ways, each of which stems from equally acceptable interpretations of Eq.~\ref{3exact}. One can equally think of Eq.~\ref{3exact} as the definition of the 3-tangle as the residual entanglement --- that which is left over from the 1-tangle and its natural decomposition into pairwise 2-tangles. One can continue to recursively define the residual entanglement for a set of $\mathcal{P}$ qubits, defined on pure states as $\tau_\text{res}=\tau_\mathcal{P} = \sqrt{\tau_A^2 - \sum_{\mathcal{I}_A<\mathcal{P}} \tau_{\mathcal{I}_A}^2}$ and extended to mixed states via convex roof. The agreement with the proposition is left to the supplementary materials \cite{ABMsup}.

Another way to interpret Eq.~\ref{3exact} is as a coincidence from taking the 3-tangle defined as an $SL$ invariant. Depending on the parity of the number of qubits, we write the $k$-tangle as the magnitude of an anti-linear operator's expectation value. For even number of pure qubits, define $\tau_\mathcal{P} = |\bra{\psi^*} \Theta^{(+)} \ket{\psi}|$, with matrix elements $\Theta^{(+)}_{i,j} = \prod_{l=0}^{k-1} \epsilon_{i_l,j_l}$ where $i_l$ is the $l$th bit of the binary expansion of $i$, $|\mathcal{P}|=k$, and $\epsilon$ is the 2-index Levi-Civita symbol \cite{LT03,WC01,GW10,GW13}. For odd number of qubits, $|\bra{\psi^*} \Theta^{(+)} \ket{\psi}|=0$ on all states and we actually must use at least a degree-4 invariant \cite{GW13}. In that case, we will write the tangle as an expectation value of an operator with the state embedded into a space of squared size, $\sqrt{2 |\bra{\psi^*} \bra{\psi^*} \Theta^{(-)} \ket{\psi} \ket{\psi}|}$, where the matrix elements are given as $\Theta^{(-)}_{i,j} = \epsilon_{i_0,i_{k}}\epsilon_{j_0,j_{k}} \prod_{l=1}^{k-1} \epsilon_{i_l,j_l} \epsilon_{i_{k+l},j_{k+l}}$. Both even/odd $k$-tangles naturally generalize the 2- and 3-tangle, and due to the anti-linearity, the convex roof can be evaluated on mixed states \cite{U}, but the explicit agreement with the proposition is left to the supplementary materials. It is worth pointing out that these tangles generalize other properties of the 2- and 3-tangle, for example for arbitrary $\mathcal{I}$ odd qubits,
\begin{equation}
\tau_{\mathcal{I}}^2= \hat{\tau}_{\mathcal{I}\setminus\{ A\} }^2- \check{\tau}_{\mathcal{I}\setminus\{ A\} }^2,
\end{equation}
generalizing Eq.~\ref{tauassistance}, recycling the notation of hat for concave roof and check for convex roof; see the supplementary materials for the proof \cite{ABMsup}. The above two interpretations of the $k$-tangle are not expected to be the same in general, so it is further surprising that they agree at all on the GHZ class.

Since strong monogamy is known to not hold in general \cite{ROA16}, it is quite curious that Eq.~\ref{mainresult} holds at all. The equation is a relation among low degree polynomial invariants, so we suppose that this special property may be a consequence of algebraically independent high degree invariants vanishing on the GHZ state. Such a guess is supported in four qubits, as the fundamental $SL$ invariants evaluate on the GHZ state as $(H^{(2)},M^{(4)}, L^{(4)}, D_{xt}^{(6)}) = (1,0,0,0)$, using the notation in \cite{LT03}, with the degrees written in the exponents.

Rather than try to rehabilitate the globally broken strong monogamy, let us paint a different picture. The take-home lesson that monogamy taught us is that the tensor product structure limits the way in which a multipartite wave function can be compatible with the tensor factors of Hilbert space. Compatibility can be well-captured by inequalities on polynomial entanglement measures. Consider a space whose dimensions are placeholders for state coefficients of a multipartite wavefunction (thought of as a real space of twice the dimension) in addition to some number of polynomial entanglement invariants, $\{ (\psi_i , I_j )\}$. An algebraic variety is defined within this space by the relationships between the invariants and the state coefficients $\{ (\psi_i , I_j ) | f_k(\psi_i,I_j) = 0\}$, where the $f$s are algebraic functions. We can project out all of the state variables from the variety to get exact constraints on the allowed entanglement invariants. Invoking the Tarski-Seidenberg theorem \cite{BPF}, which states the projection of a semi-algebraic set is again a semi-algebraic set, then proves that the achievable set of any collection of polynomial entanglement measures (a projection) is described by a collection of semi-algebraic relations on those measures. The advantage of this framework is that the resulting inequalities are not only necessary, but also sufficient.

One algorithm in particular is known for performing such projections, the cylindrical decomposition, however it is in general a doubly exponential algorithm in the number of dimensions \cite{BPF}. The 3-qubit case is the simplest system where non-trivial entanglement trade-off appears, yet it seems inefficient for the algorithm to handle blindly. Therefore, we have simplified the problem, to, in a sense, manually perform the projection. We have then given a few concrete examples of the Tarski-Seidenberg theorem in action (some other examples can be found in \cite{AM17, BGM, BDM}) which in our case has led us to even stronger constraints beyond monogamy. Inasmuch as the GHZ state is an economically viable reagent in quantum experimentation and can provide an operational meaning to the $k$-tangle, a comprehensible theory of multi-partite entanglement may be a step closer to being within grasp. On the other hand, it is known that the marginal inequalities become drastically complicated \cite{K04}, while still remaining linear in eigenvalues, and we thus might expect semi-algebraic relations on invariants, which apparently imply marginal constraints, to become even more vastly complicated.



\bibliographystyle{h-physrev}
\bibliography{refs}

\widetext
\clearpage
\begin{center}
\textbf{\large Supplemental Materials: Entanglement Constraints on States Locally Connected to the Greenberger-Horne-Zeilinger State}
\end{center}
\setcounter{equation}{0}
\setcounter{figure}{0}
\setcounter{table}{0}
\setcounter{page}{1}
\makeatletter
\renewcommand{\theequation}{S\arabic{equation}}
\renewcommand{\thefigure}{S\arabic{figure}}


\section{3-qubit Tangle Parametrization}
Mixed state 2-tangles have the following general formula, $\tau_{A|B}(\rho_{AB}) = \text{Max}(\sqrt{\lambda_1} - \sqrt{\lambda_2} - \sqrt{\lambda_3} - \sqrt{\lambda_4},0)$, where the $\lambda_i$s are the non-ascending eigenvalues of the operator $R=\rho_{AB} (\sigma_y\otimes \sigma_y) \rho_{AB}^{T}(\sigma_y \otimes \sigma_y)$ \cite{W98}.
The 2-party marginals of the unnormalized $\ket{\psi_{ABC}}=M_1\otimes M_2\otimes M_3 \ket{\text{GHZ}}$ are rank-2, and the 2-tangle can be quickly computed with the simplified formula, 
\begin{align}
\begin{split}\label{rank2tau}
p \text{\space}\tau_{A|B} &= \sqrt{ \lambda_1} - \sqrt{\lambda_2} \\
		&= \sqrt{ \lambda_1 + \lambda_2 - 2\sqrt{\lambda_1 \lambda_2}} \\
		&= \sqrt{\text{Tr}(R)-\sqrt{2( \text{Tr}(R)^2-\text{Tr}(R^2)) }},
\end{split}
\end{align}
with the now necessary normalization factor $p = \braket{\psi_{ABC}|\psi_{ABC}}$. If we parametrize the $M_i$s with two 2-component complex column vectors $M_i = ( \underline{u_i} \text{\space } \underline{v_i})$, where underscore means the coordinates of a geometric vector, then the resulting 2-tangle expression simplifies to $\tau_{A|B}  =\frac{1}{p} |\vec{u_1} \wedge \vec{v_1}|  |\vec{u_2} \wedge \vec{v_2}|  |\vec{u_3}^\dagger  \vec{v_3}|$, with $p =\frac{1}{2} |\vec{u_1} \vec{u_2} \vec{u_3} + \vec{v_1} \vec{v_2} \vec{v_3}|^2$ and the vector juxtaposition means tensor product.

Due to the form of the expression with absolute values, complex numbers in the matrices $M_i$ become redundant, and one can achieve the same values of the tangles with real matrices. Therefore it is convenient to parametrize the columns as $\underline{u_i} = (u_i \cos(\theta_i) \text{\space} u_i \sin(\theta_i))^T$ and $\underline{v_i} = (v_i \cos(\theta_i+\phi_i) \text{\space} v_i \sin(\theta_i+\phi_i))^T$, whence the tangle simplifies to $\tau_{A|B}  =\frac{2}{p} u_1 u_2 u_3 v_1 v_2 v_3 | s_1 s_2 c_3 |$, where $s_i = \sin{\phi_i}$ and $c_i = \cos{\phi_i}$, and $p = u_1^2 u_2^2 u_3^2 + v_1^2 v_2^2 v_3^2 + 2u_1 u_2 u_3 v_1 v_2 v_3 c_1 c_2 c_3$. Notice that the $\theta_i$s do not appear in the expressions so we could set them to zero.  Simplifying further, we get $\tau_{A|B}  =| s_1 s_2 c_3 | / (r + c_1 c_2 c_3)$, where $2r = u_1 u_2 u_3 / v_1 v_2 v_3 + v_1 v_2 v_3 / u_1 u_2 u_3$ and since all the $u_i, v_i \geq0$, we have the bound $r\geq 1$ and consequentially, we only need to consider the case where $c_i \leq 0$. Accordingly, we shall make the minus sign on the $c_i$s explicit and then only consider angles $\phi_i \in [0,\pi/2]$, which allows us to remove the absolute value signs. All of the 2-tangles can be cleansed of redundancy in an analogous manner, which then amounts to permuting the subscripts. 

The 3-tangle can be calculated from the general formula \cite{CKW00},
\begin{equation}
\tau_{A|B|C}(\psi) = \sqrt{2 | \psi_{j_5,j_4,j_3}\psi_{j_2,j_1,j_0} \epsilon_{i_0, i_3}\epsilon_{j_0, j_3}\epsilon_{i_1, j_1}\epsilon_{i_2, j_2}\epsilon_{i_4, j_4}\epsilon_{i_5, j_5}  \psi_{i_5,i_4,i_3}\psi_{i_2,i_1,i_0}|},
\end{equation}
with $\epsilon$ being the 2-index Levi-Civita symbol. It is easy to check that $\tau_{A|B|C}(\ket{\text{GHZ}}) = 1$, and then apply the transformation rule, $\frac{1}{p}\tau(\otimes_i M_i \ket{\text{GHZ}})  = \frac{1}{p}\prod_i | \text{det}(M_i)| \tau(\ket{\text{GHZ}}) = \frac{1}{p}  |\vec{u_1} \wedge \vec{v_1}|  |\vec{u_2} \wedge \vec{v_2}|  |\vec{u_3} \wedge  \vec{v_3}|$ with $p$ as before, which can then be reparametrized in the same manner as the 2-tangles. Thus we reproduce the expressions from the main text, repeated here for convenience,
\begin{align}
\begin{split}
\label{Sxyzt}
x &= c_1 s_2 s_3/ (r- c_1 c_2 c_3), \text{\space \space}y  =  s_1 c_2 s_3 / (r- c_1 c_2 c_3), \\
z&=  s_1 s_2 c_3/ (r- c_1 c_2 c_3), \text{\space \space } t=  s_1 s_2 s_3 / (r- c_1 c_2 c_3),
\end{split}
\end{align}
where $(\tau_{B|C}, \tau_{A|C}, \tau_{A|B}, \tau_{A|B|C}) \equiv (x,y,z,t)$.

\section{Proof of Achievability Theorem}

To prove the theorem, it will be useful to invert the tangle expressions, Eq.~\ref{Sxyzt}, for the parameters. A Gr\"obner elimination based inversion must proceed in two calculations due to the degenerate case when $d = 2(r - c_1 c_2 c_3) = 0$. Therefore we find the Gr\"obner basis with the above equations, eliminating the angles, but keeping $(x,y,z,t,d)$, to give the relation,
\begin{equation}
d \left(d^2 (t^2 + x^2) (t^2 + y^2) (t^2 + z^2) - 4 t^4 \right) = 0.
\end{equation}
We drop the degeneracy causing factor of $d=0$, and add the remaining factor back to the same set of equations above. Computing another Gr\"obner basis for each variable, eliminating two angles each time as well as $d$, gives the following inversion,
\begin{equation}
\label{cccr}
c_1 =  \frac{x}{\sqrt{t^2 +x^2}}, \text{\space \space} c_2 =  \frac{y}{\sqrt{t^2 +y^2}},   \text{\space \space} c_3 =  \frac{z}{\sqrt{t^2 +z^2}},
\end{equation}
which can then be back-substituted to find $r$ as well,
\begin{equation}
r = \frac{t^2+x y z }{\sqrt{ (t^2 + x^2) (t^2 + y^2) (t^2 + z^2)}}.
\end{equation}
The only non-trivial bound is $r\geq1$, where by expanding the expression gives the inequality of the theorem, thus explicitly proving sufficiency. 

To show necessity of the inequality, we just plug in expressions from Eq.~\ref{Sxyzt} into the Eq.~\ref{xyztineq} and it can be simplified to,
\begin{equation}
\frac{s_1^2 s_2^2 s_3^2 (r^2-1)}{(r-c_1 c_2 c_3)^4} \geq 0,
\end{equation}
and since $r\geq 1$, the inequality is true.

The above argument is only for \textit{generic} states, and rather than appealing to continuity we'd like to provide alternative evidence that the inequality is true for \textit{arbitrary} states. It will be convenient to use expressions for $(x,y,z,t)$ from the 3-qubit Schmidt form \cite{AAJT,ES14}, using the unitary invariance of the invariants to simplify an arbitrary state,
\begin{equation}\label{canonicalform}
\psi = (a_1,a_2,a_3,a_4,a_5,a_6,a_6,a_7,a_8) \xrightarrow{U_A \otimes U_B \otimes U_C} (\lambda_0,0,0,0,\lambda_1 e^{i \omega},\lambda_2,\lambda_3,\lambda_4),
\end{equation}
with real parameters $(\lambda_0,\lambda_1, \lambda_2, \lambda_2, \lambda_4,\omega)\geq 0$. The invariants can be computed straight forwardly \textit{\`a la} Eq.~\ref{rank2tau},
\begin{align}
\begin{split}
\label{acin}
x &= 2 |\lambda_2 \lambda_3 - e^{i \omega} \lambda_1 \lambda_4|,  \text{ \space \space } y = 2 \lambda_0 \lambda_2, \text{ \space \space } z = 2 \lambda_0 \lambda_3, \text{ \space \space } t = 2 \lambda_0 \lambda_4,
\end{split}
\end{align}
where again, $(\tau_{B|C}, \tau_{A|C}, \tau_{A|B}, \tau_{A|B|C}) \equiv (x,y,z,t)$. Notice that $\omega$ only varies $x$ independently of the other invariants. The left-hand side of Eq~\ref{xyztineq}, turns out to be concave in $x$, as seen by taking two derivatives,
\begin{equation}
\partial_x^2 [t^2 ( 1-x^2-y^2-z^2-t^2)-( x^2 y^2 + x^2 z^2 +y^2 z^2 - 2 x y z) ] = -2 (t^2 + y^2 + z^2),
\end{equation}
so we only need to check for non-negativity with the extreme values of $x$, meaning $\omega = 0, \text{\space} \pi$. Plugging in expressions, Eq.~\ref{acin} with the case of $\omega=\pi$, that is, $x= 2 (\lambda_2 \lambda_3 +\lambda_1 \lambda_4)$, turns the lhs of the inequality into,
\begin{align}
\begin{split}
(\lambda_0^2+\lambda_1^2+\lambda_2^2+\lambda_3^2+\lambda_4^2-1) g + [2 \lambda_0 (2 \lambda_1 \lambda_2 \lambda_3 + \lambda_4 - 2 \lambda_4 (\lambda_2^2 + \lambda_3^2 + \lambda_4^2))]^2,
\end{split}
\end{align}
where $g$ is some polynomial in the $\lambda_i$s, but by applying normalization, $\lambda_0^2+\lambda_1^2+\lambda_2^2+\lambda_3^2+\lambda_4^2=1$, the first term vanishes regardless of the sign of $g$, and the remaining term is a perfect square, and hence non-negative certified. Now plugging in expressions for $(x,y,z,t)$ in the other case of  $\omega=0$, that is, $x= 2 (\lambda_2 \lambda_3 -\lambda_1 \lambda_4)$, assuming for now that $\lambda_2 \lambda_3 \geq \lambda_1 \lambda_4$, turns the lhs of the inequality into,
\begin{align}
\begin{split}
(\lambda_0^2+\lambda_1^2+\lambda_2^2+\lambda_3^2+\lambda_4^2-1) g' + [2 \lambda_0 (-2 \lambda_1 \lambda_2 \lambda_3 + \lambda_4 - 2 \lambda_4 (\lambda_2^2 + \lambda_3^2 + \lambda_4^2))]^2,
\end{split}
\end{align}
with a new polynomial $g'$ --- amounting to a very minor change, but which importantly preserves the existence of a perfect square non-negative certificate. If we assume that $\lambda_2 \lambda_3 \leq \lambda_1 \lambda_4$, then the expressions of $(x,y,z,t)$, with $x= -2 (\lambda_2 \lambda_3 -\lambda_1 \lambda_4)$, obviously satisfy, 
\begin{equation}
t^2 ( 1-x^2-y^2-z^2-t^2)-( x^2 y^2 + x^2 z^2 +y^2 z^2 + 2 x y z) \geq 0,
\end{equation}
where the only change from the theorem's inequality is a minus sign on the $x$ with a unit exponent. The above constraint is actually a stronger constraint than the theorem's inequality and therefore the theorem is proved.

\section{Univariate Marginal Eigenvalue Inequality}

Take the set of relations of eigenvalues and invariants,
\begin{align}
\begin{split}
\lambda_A &= \frac{1}{2}\left( 1-\sqrt{1-\tau_{A|B}^2 - \tau_{A|C}^2-\tau_{A|B|C}^2} \right),
\end{split}
\end{align}
where other relations are obtained by permuting the parties and each $\lambda_P$ is the smallest eigenvalue of party $P$. Inverting, we get,
\begin{align}
\begin{split}
\tau_{A|B}=&\sqrt{2\lambda_A(1-\lambda_A)+2\lambda_B (1-\lambda_B)-2\lambda_C (1-\lambda_C)-\tau_{A|B|C}^2/2},
\end{split}
\end{align}
where again, other relations are obtained by permuting the parties. We substitute into the inequality of theorem to get the following,

\begin{align} \label{lamtau}
\begin{split}
& \tau_{A|B|C}^2 - \tau_{A|B|C}^4/4+ 4(\lambda_A^4+\lambda_B^4+\lambda_C^4)-8(\lambda_A^3+\lambda_B^3+\lambda_C^3)+4(\lambda_A^2+\lambda_B^2+\lambda_C^2)\\
&\quad +8\left( \lambda_A(1-\lambda_A)\lambda_B(1-\lambda_B)+\lambda_A(1-\lambda_A)\lambda_C(1-\lambda_C)+\lambda_B(1-\lambda_B)\lambda_C (1-\lambda_C)\right)\\
&\quad+2\sqrt{2\lambda_A(1-\lambda_A)+2\lambda_B (1-\lambda_B)-2\lambda_C (1-\lambda_C)-\tau_{A|B|C}^2/2} \\
&\quad\quad\times  \sqrt{2\lambda_A(1-\lambda_A)-2\lambda_B (1-\lambda_B)+2\lambda_C (1-\lambda_C)-\tau_{A|B|C}^2/2}\\
&\quad\quad\times  \sqrt{-2\lambda_A(1-\lambda_A)+2\lambda_B (1-\lambda_B)+2\lambda_C (1-\lambda_C)-\tau_{A|B|C}^2/2} \geq 0.
\end{split}
\end{align}
Since it is easier to work with polynomials, we will take the above inequality, turn it into an equality and square away the root. The resulting expression has two factors, $p_1 p_2 =0$, where,
\begin{align*}
p_1&=\tau_{A|B|C}^4 + 16 (1+\lambda_A -\lambda_B-\lambda_C)(1-\lambda_A+\lambda_B-\lambda_C)(1-\lambda_A-\lambda_B+\lambda_C)(1-\lambda_A-\lambda_B-\lambda_C),\\
p_2&=\tau_{A|B|C}^4+16(\lambda_A-\lambda_B-\lambda_C)(-\lambda_A+\lambda_B-\lambda_C)(-\lambda_A-\lambda_C+\lambda_C)(2-\lambda_A-\lambda_B-\lambda_C).
\end{align*}
Each factor can be thought of as the boundary of the maximal eigenvalues, or the minimal eigenvalues of each party since the factors are related by a substitution, $\lambda_P \mapsto 1-\lambda_P$. Of the two factors, the latter will determine the constraint on the minimal eigenvalues. One can again simply set $\tau_{A|B|C}=0$, to maximally unconstrain the eigenvalues, and it is then valid to drop the last factor, $(2-\lambda_A-\lambda_B-\lambda_C)$, since it has no bearing on the minimal eigenvalues. Thus we recover the marginal eigenvalue bound.

\section{k-tangle Parametrization}

The beginning notions of strong monogamy first considered $k$-tangles as \textit{residual} tangles \cite{RLMA}, defined on pure states as $\tau_\text{res}=\tau_\mathcal{P} = \sqrt{\tau_A^2 - \sum_{\mathcal{I}_A<\mathcal{P}} \tau_{\mathcal{I}_A}^2}$ and extended to mixed states via convex roof. First define $\ket{\text{GHZ}}_{a,b} = a \ket{0}^{\otimes n} + b \ket{1}^{\otimes n}$, and since a loss of any qubit results in a separable state, we have $\tau_\text{res}(\ket{\text{GHZ}}_{a,b}) = \tau_A(\ket{\text{GHZ}}_{a,b}) = 2 |ab|$ for any party $A$. Notice the \textit{same} formula results from the 2-tangles on the same pure 2-qubit GHZ state. We can therefore use the 2-tangle mixed state formula to compute the residual tangle on a mixture of two $n$-qubit $\ket{\text{GHZ}}_{a,b}$ states with differing $a,b$, since \textit{every} corresponding pure state decomposition will be a mixture of only GHZ states (where the minimum is already known by the 2-tangle formula). A mixture of two GHZ states forms a matrix which has its only non-zero block on the $\ket{0}^{\otimes n}, \ket{1}^{\otimes n}$ subspace --- the tangle computation gives, $\tau_\text{res} \begin{pmatrix} \alpha & \beta \\ \bar{\beta} & \gamma \end{pmatrix}   = 2 |\beta|$. The result can be applied to the following subsystem, for general $M_p = ( \underline{u_p}, \text{\space} \underline{v_p}), \vec{u}_p,\vec{v}_p \in \mathbb{C}^{\otimes 2}$,
\begin{align}
\begin{split}\label{bigrho}
\rho_\mathcal{I} &=\text{Tr}_{\bar{\mathcal{I}}} (\otimes_{\bar{\imath} \in \bar{ \mathcal{I} }} M_{\bar{\imath}}  \ket{\text{GHZ}}) \\
&= \begin{pmatrix} \prod_{\bar{\imath}} \vec{u}_{\bar{\imath}}^\dagger \vec{u}_{\bar{\imath}} & \prod_{\bar{\imath}} \vec{u}_{\bar{\imath}}^\dagger \vec{v}_{\bar{\imath}}  \\  \prod_{\bar{\imath}} \vec{v}_{\bar{\imath}}^\dagger \vec{u}_{\bar{\imath}} &  \prod_{\bar{\imath}} \vec{v}_{\bar{\imath}}^\dagger \vec{v}_{\bar{\imath}}  \end{pmatrix},
\end{split}
\end{align}
where what is shown is the only non-zero block which has its support in the $\ket{0}^{\otimes n}, \ket{1}^{\otimes n}$ subspace. Notice that the $M_{\bar{\imath}}$s here are acting externally to $\mathcal{I}$. These residual tangles can be shown inductively to transform under internal SLOCC on the pure GHZ class as $\tau_{\text{res}}(\otimes_k M_k \ket{\text{GHZ}}) = \prod_k |\text{det}(M_k)| \tau_{\text{res}} (\ket{\text{GHZ}})$, and if the rule holds on pure states, it also holds on mixed states. The base case of the 2-tangles is already known to obey the transformation rules on pure and mixed states. Assuming the transformation rules hold for the $k$-tangles, $k<n$, we can assume the validity of expressions in the proposition (except the $n$-tangle), but Eq.~\ref{mainresult} then defines the $n$-tangle, which thus gives the formula in the proposition as well as the transformation rule on pure states. The extension of the transformation rule to mixed states is a result of Tajima \cite{T13}. So as long as an $M_i$ isn't traced over, it can be factored out of the tangle as a determinant. Thus when all is fully evaluated, we reproduce the tangle expressions from the proposition.

The previous interpretation of $k$-tangles as residual tangles has the sense of a tautology, since Eq.~\ref{mainresult} is, after all, the definition of the $n$-qubit $n$-tangle as a residual tangle. Just as in the 3-qubit case, there is an independent interpretation of $k$-tangles which arrives at the same, Eq.~\ref{mainresult}, in a highly non-obvious way. For even $k$, the pure state $k$-tangle can be defined as $|\bra{\psi^*} \Theta^{(+)} \ket{\psi}|$, with matrix elements $\Theta^{(+)}_{i,j} = \prod_{l=0}^{k-1} \epsilon_{i_l,j_l}$  where $i_l$ is the $l$th bit of the binary expansion of $i$, which is known as Caley's \textit{other} hyperdeterminant, and has been considered by several others \cite{LT03,WC01,GW10,GW13}. Due to the anti-linear hermiticity, a complete formula for the convex roof is given in \cite{U} as a generalization of Wootter's formula,
\begin{equation}
\tau_{\mathcal{I}}(\rho_\mathcal{I}) = \text{Max}(\sqrt{\lambda_1} - \sum_{i\geq2} \sqrt{\lambda_i},0),
\end{equation}
with $\lambda_i$s the non-ascending eigenvalues of $R=\rho_\mathcal{I} \Theta^{(+)} \rho_\mathcal{I} \Theta^{(+)}.$ The tangle transforms as usual under internal local $GL$ operations, and a direct calculation with Eq~\ref{bigrho} and Eq.~\ref{rank2tau} readily recovers the expressions in the proposition.

As mentioned in the main text, for odd $k$, $\Theta^{(+)}$ vanishes on all states and we are forced to use at least a degree-4 polynomial \cite{GW13}. In that case, we will write the tangle as an expectation value of an operator with the state embedded into a space of twice the size, $\sqrt{2 |\bra{\psi^*} \bra{\psi^*} \Theta^{(-)} \ket{\psi} \ket{\psi}|}$, where the matrix elements are given as $\Theta^{(-)}_{i,j} = \epsilon_{i_0,i_{k}}\epsilon_{j_0,j_{k}} \prod_{l=1}^{k-1} \epsilon_{i_l,j_l} \epsilon_{i_{k+l},j_{k+l}}$, which again has an analogous formula for the convex roof \cite{U}. There is an ambiguity in how to embed an odd $k$-qubit mixed state into $2k$-qubit space, we propose the option, $\rho \mapsto \rho \otimes \rho$. With this definition of the odd $k$-tangle, the formulas in the proposition can be reproduced.

\section{k to k-1 tangle relation}

For an arbitrary set of odd $k$ pure qubits, $\mathcal{I}$, not necessarily in the GHZ class, the $k$-tangle satisfies an analogous property as the 2- and 3-tangle. Consider the definition,
\begin{align}
\begin{split}
\tau_{\mathcal{I}}^2 &= 2 |\bra{\psi^*} \bra{\psi^*} \Theta^{(-)} \ket{\psi} \ket{\psi}| \\
&= 2\sqrt{\bra{\psi^*} \bra{\psi^*} \Theta^{(-)} \ket{\psi} \ket{\psi} \bra{\psi} \bra{\psi} \Theta^{(-)} \ket{\psi^*} \ket{\psi^*}},
\end{split}
\end{align}
and write out the term under the root in components as,
\begin{align}
\begin{split}\label{components}
\psi_{i_0, i_<}\psi_{i_k,i_>}\psi_{j_0, j_<}\psi_{j_k,j_>} \psi_{a_0, a_<}^*\psi_{a_k, a_>}^*\psi_{b_0, b_<}^*\psi_{b_k,b_>}^* \epsilon_{i_0,i_{k}}\epsilon_{j_0,j_{k}} \epsilon_{a_0,a_{k}}\epsilon_{b_0,b_{k}}\epsilon_{i,j}^< \epsilon_{i,j}^> \epsilon_{a,b}^< \epsilon_{a,b}^>,
\end{split}
\end{align}
where we use compact notation to save space, $i_< =  i_1,\ldots,i_{k-1}$,  $i_> = i_{k+1},\ldots,i_{2k-1}$ and $\epsilon_{i,j}^< = \prod_{l=1}^{k-1} \epsilon_{i_l,j_l}$,  $\epsilon_{i,j}^> = \prod_{l=1}^{k-1} \epsilon_{i_{k+l},j_{k+l}}$. Note, we have specifically shunned the 0-bit and $k$-bit indices from the compacted notation for isolated computations. By applying the identity, $\epsilon_{i,i'}\epsilon_{j,j'} = \delta_{i,j}\delta_{i',j'} - \delta_{i,j'}\delta_{i',j}$, between $i$s and $a$s, and again between $j$s and $b$s, we get,
\begin{align}
\begin{split}
\epsilon_{i_0,i_{k}}\epsilon_{j_0,j_{k}} \epsilon_{a_0,a_{k}}\epsilon_{b_0,b_{k}} &= ( \delta_{i_0,a_0}\delta_{i_k,a_k} - \delta_{i_0,a_k}\delta_{i_k,a_0}) ( \delta_{j_0,b_0}\delta_{j_k,b_k} - \delta_{j_0,b_k}\delta_{j_k,b_0}) \\
&=  \delta_{i_0,a_0}\delta_{i_k,a_k}\delta_{j_0,b_0}\delta_{j_k,b_k}  -  \delta_{i_0,a_0}\delta_{i_k,a_k} \delta_{j_0,b_k}\delta_{j_k,b_0} -\delta_{i_0,a_k}\delta_{i_k,a_0}\delta_{j_0,b_0}\delta_{j_k,b_k} +  \delta_{i_0,a_k}\delta_{i_k,a_0}\delta_{j_0,b_k}\delta_{j_k,b_0}.
\end{split}
\end{align}
Now, we simplify each term, one at a time within Eq.~\ref{components}. The first term gives,
\begin{align}
\begin{split}\label{term1}
&\psi_{i_0, i_<}\psi_{i_k,i_>}\psi_{j_0, j_<}\psi_{j_k,j_>} \psi_{a_0, a_<}^*\psi_{a_k, a_>}^*\psi_{b_0, b_<}^*\psi_{b_k,b_>}^* \delta_{i_0,a_0}\delta_{i_k,a_k}\delta_{j_0,b_0}\delta_{j_k,b_k} \epsilon_{i,j}^< \epsilon_{i,j}^> \epsilon_{a,b}^< \epsilon_{a,b}^> \\
&= \rho_{i_<,a_<} \rho_{i_>,a_>} \rho_{j_<,b_<} \rho_{j_>,b_>}\epsilon_{i,j}^< \epsilon_{i,j}^> \epsilon_{a,b}^< \epsilon_{a,b}^> \\
&= (\rho_{i_<,a_<} \epsilon_{a,b}^< \rho^T_{b_<,j_<} \epsilon_{j,i}^<) (\rho_{i_>,a_>}  \epsilon_{a,b}^> \rho^T_{b_>,j_>} \epsilon_{j,i}^> ) \\
&=  \text{Tr}(R)^2,
\end{split}
\end{align}
with $\rho_{i_<,a_<} = \psi_{i_0, i_<}\psi_{a_0,a_<}^* \delta_{i_0,a_0}$ being the components of $\rho_{\mathcal{I}\setminus\{ A\}} = \text{Tr}_{A} \ket{\psi_\mathcal{I}} \bra{\psi_\mathcal{I}}$, and $R = \rho_{\mathcal{I}\setminus\{ A\}} \Theta^{(+)} \rho_{\mathcal{I}\setminus\{ A\}}^{T} \Theta^{(+)}$.
The second term gives,
\begin{align}
\begin{split}\label{term1}
&\psi_{i_0, i_<}\psi_{i_k,i_>}\psi_{j_0, j_<}\psi_{j_k,j_>} \psi_{a_0, a_<}^*\psi_{a_k, a_>}^*\psi_{b_0, b_<}^*\psi_{b_k,b_>}^* \delta_{i_0,a_0}\delta_{i_k,a_k} \delta_{j_0,b_k}\delta_{j_k,b_0} \epsilon_{i,j}^< \epsilon_{i,j}^> \epsilon_{a,b}^< \epsilon_{a,b}^> \\
&= \rho_{i_<,a_<} \rho_{i_>,a_>} \rho_{j_<,b_>} \rho_{j_>,b_<}\epsilon_{i,j}^< \epsilon_{i,j}^> \epsilon_{a,b}^< \epsilon_{a,b}^> \\
&= (\rho_{i_<,a_<} \epsilon_{a,b}^< \rho^T_{b_<,j_>} \epsilon_{j,i}^>) (\rho_{i_>,a_>}  \epsilon_{a,b}^> \rho^T_{b_>,j_<} \epsilon_{j,i}^< ) \\
&=  \text{Tr}(R^2),
\end{split}
\end{align}
using the same notation as before. The third and fourth term calculations recapitulate the first and second, and therefore we have the following,
\begin{align}
\begin{split}
\tau_{\mathcal{I}}^2 &=2\sqrt{2( \text{Tr}(R)^2-\text{Tr}(R^2)) }\\
&= \text{Tr}(R)+\sqrt{2( \text{Tr}(R)^2-\text{Tr}(R^2))} \\ 
   & \text{\space \space\space\space}-\left(\text{Tr}(R) - \sqrt{2( \text{Tr}(R)^2-\text{Tr}(R^2))} \right) \\
   &= (\sqrt{\lambda_1} + \sqrt{\lambda_2})^2 - (\sqrt{\lambda_1} - \sqrt{\lambda_2})^2 \\
&= \hat{\tau}_{\mathcal{I}\setminus\{ A\} }^2- \check{\tau}_{\mathcal{I}\setminus\{ A\} }^2,
\end{split}
\end{align}
so that these generalized $k$-tangles maintain similar properties of the few-party tangles as one might expect, recall Eq~\ref{rank2tau} and Eq~\ref{tauassistance}.

\end{document}